\documentclass[11pt]{article}%
\usepackage{amsfonts}
\usepackage{amsmath}
\usepackage{amssymb}
\usepackage{graphicx}%
\providecommand{\U}[1]{\protect\rule{.1in}{.1in}}
\newtheorem{theorem}{Theorem}
\newtheorem{acknowledgement}[theorem]{Acknowledgement}

\begin{document}

\title{Reconstruction of Gaussian Quantum States from Ideal Position Measurements:
Beyond Pauli's Problem, I}
\author{Maurice A. de Gosson\thanks{maurice.de.gosson@univie.ac.at}\\University of Vienna\\Faculty of Mathematics (NuHAG)\\Oskar-Morgenstern-Platz 1\\1090 Vienna AUSTRIA}
\maketitle

\begin{abstract}
We show that the covariance matrix of a quantum state can be reconstructed
from position measurements using the simple notion of polar duality, familiar
from convex geometry. In particular, all multidimensional Gaussian states
(pure or mixed) can in principle be reconstructed if the quantum system is
well localized in configuration space. The main observation which makes this
possible is that the John ellipsoid of the Cartesian product of the position
localization by its polar dual contains a quantum blob, and can therefore be
identified with the covariance ellipsoid of a quantum state.

\end{abstract}

\section{Introduction}

The notion of a point particle is an idealization heavily used in physics. We
have introduced in a recent preprint \cite{MC} the more realistic notion of
\textquotedblleft pointillism\textquotedblright\ which associates to a set of
position observations of the particle a convex set $X$ in configuration space
(for instance an ellipsoid) supposed to represent physically that particle
(whether it has an internal structure or not); this is consistent with the
fact that the particle's wavepacket always occupies a nonzero volume. We have
then postulated in \cite{MC} that the momentum of the particle is represented
by a (shift of) what we have called the $\hbar$-polar dual set $X^{\hbar}$ of
$X$. Suppose that $X$ is an ellipsoid centered at the origin in configuration
space and that the momenta are located around the origin in momentum space.
Then, by definition, $X^{\hbar}$ consists of all momentum vectors
$p=(p_{1},...,p_{n})$ such that
\begin{equation}
p_{1}x_{1}+\cdot\cdot\cdot+p_{n}x_{n}\leq\hbar\label{polar}%
\end{equation}
for all position vectors $x=(x_{1},...,x_{n})$ in $X$. This assumption is
consistent with the uncertainty principle for position and momentum which
roughly states that one cannot assign exact simultaneous values to the
position and momentum of a physical system \cite{Huff3}. Suppose, for example,
that we are studying the atomic orbit of an electron in the hydrogen atom; it
occupies a volume $\Delta V\sim10^{-30}m^{3}$ which we model a a sphere with
radius roughly $\Delta r\sim10^{-10}m$. The $\hbar$-polar dual of such a
sphere is a sphere in momentum space with radius $\Delta p=10^{10}\cdot\hbar$
so that $\Delta r\Delta p\sim\hbar$. We are going to show in this Letter that
a slight generalization of the notion of $\hbar$-polar duality allows the
construction of all Gaussian states (or, more generally, covariance matrices)
corresponding to a given set of ideal position measures.

\paragraph{On the use of standard deviations in quantum mechanics}

As we already have explained in \cite{MPAG}, the usual description of states
using standard deviations (variances and covariances) is \textit{ad hoc} and
not always good choice to quantify uncertainties; Uffink and Hilgevoord have
analyzed this in a series of works \cite{Huff0,Huff1,Huff2,Huff3} where they
show that standard deviation is not an adequate measure of quantum uncertainty
in general, and only gives an acceptable measurement of the spread of a
wavefunction when the probability density is Gaussian, or nearly Gaussian. In
the present paper we do not make any \textit{a prior}i assumption about the
existence of any particular probability distribution in the measurements, but
only assume that these are well localized (contained in ellipses or ellipsoids).

\paragraph{The Pauli reconstruction problem}

Our constructions (which complete and extend those we gave in \cite{gopauli})
are related to Pauli's reconstruction problem: is it possible to determine a
state by knowing only the configuration space and the momentum space marginal
distributions? The answer is known to be negative: in general the knowledge of
position and momentum marginals leads to several states (\textquotedblleft
Pauli partners\textquotedblright); for a simple example involving Gaussians
see \cite{espo}. For details see for instance \cite{Weigert} which contains an
interesting discussion and many references to the Pauli problem. A fact to be
noticed (and which is not usually taken into account in the literature) is
that the ambiguity of the reconstruction problem comes from the fact that in
the general case the knowledge of the standard deviations $\sigma_{x_{j}x_{j}%
}$ and $\sigma_{p_{j}p_{j}}$ in the position and momentum variables do not
suffice to determine the correlations (covariance) $\sigma_{x_{j}p_{j}}$: the
Robertson--Schr\"{o}dinger inequalities
\begin{equation}
\sigma_{x_{j}x_{j}}\sigma_{p_{j}p_{j}}-\sigma_{x_{j}p_{j}}^{2}\geq\tfrac{1}%
{4}\hbar^{2}%
\end{equation}
do not allow to determine $\sigma_{x_{j}p_{j}}^{2}$ --let alone $\sigma
_{x_{j}p_{j}}$--. An exception is however the case of Gaussian (pure) states,
because these are known to saturate the Robertson--Schr\"{o}dinger
inequalities, that is they become the equalities%
\begin{equation}
\sigma_{x_{j}x_{j}}\sigma_{p_{j}p_{j}}-\sigma_{x_{j}p_{j}}^{2}=\tfrac{1}%
{4}\hbar^{2}%
\end{equation}
which yields to values $\pm\sigma_{x_{j}p_{j}}$ in each degree of freedom. A
\textit{caveat}: the converse is not true, because saturation of the
Robertson--Schr\"{o}dinger inequalities does not imply that the state under
scrutiny is Gaussian (see our paper \cite{digopra} with Dias and Prata for a
detailed discussion of this fact, and counterexamples).

\paragraph{Structure of the work}

This Letter is structured as follows: in Section \ref{sec1} we discuss the
elementary example of a particle constrained to move along a line; \ in
Section \ref{sec2} we discuss the multidimensional case; we first address the
case of pure states and thereafter we generalize our constructions to the case
of mixed Gaussian states. We thereafter discus the qualities and the
shortcomings of our approach, indicating possible generalizations. We have
included two Appendices: in Appendix A we review the basics of symplectic
linear algebra, and in Appendix B we give the basic definitions in the theory
of polar duality.

\section{Introductory Example\label{sec1}}

We are studying a particle moving along a line (the \textquotedblleft%
$x$-axis\textquotedblright) and we perform a set of precise position
measurements at a given time. Eliminating outliers and systemic errors by
standard statistical methods, we find the particle is located in a small
interval
\[
X=[x_{0}-\Delta x,x_{0}+\Delta x]=x_{0}+[-\Delta x,\Delta x]
\]
centered around some value $x_{0}$; we emphasize that the deviation $\Delta
x>0$ is \textit{not} at this point identified with any standard deviation
associated with a probability law. What about the momentum of the particle? We
postulate that it is located in an interval%
\[
P=[p_{0}-\Delta p,p_{0}+\Delta p]=p_{0}+[-\Delta p,\Delta p]
\]
where the interval $[-\Delta p,\Delta p]$ contains the $\hbar$-polar dual
$[-\Delta x,\Delta x]^{\hbar}$; the latter is easily seen to be the interval
$[-\hbar/\Delta x,\hbar/\Delta x]$ so that the inclusion $[-\Delta x,\Delta
x]^{\hbar}\subset\lbrack-\Delta p,\Delta p]$ implies that we must have $\Delta
p\Delta x\geq\hbar$ (but be again aware that the quantities $\Delta x$ and
$\Delta p$ are not identified with standard deviations; they are just
measurements of inaccuracy). We now ask:

\begin{quotation}
\emph{Using the information obtained by position measurements, can we
determine a plausible quantum state of the particle?}
\end{quotation}

It turns out, quite surprisingly that we can associate to the intervals
$[x_{0}-\Delta x,x_{0}+\Delta x]$ and $[p_{0}-\Delta p,p_{0}+\Delta p]$
quantum covariance matrices; if the state is known to be Gaussian this totally
determines the state in question. To simplify notation we assume that the
particle is located at the origin of the phase plane, \textit{i.e}. $x_{0}=0$,
$p_{0}=0$, the general case of non zero $x_{0}$ and $p_{0}$ is obtained using
the translations $x\longmapsto x+x_{0}$ and $p\longmapsto p+p_{0}$. Assume
first that we have the equality%
\begin{equation}
\lbrack-\Delta x,\Delta x]^{\hbar}=[-\hbar/\Delta x,\hbar/\Delta x]=[-\Delta
p,\Delta p] \label{equal1}%
\end{equation}
which is equivalent to the condition $\Delta p\Delta x=\hbar$ and consider now
the rectangle $[-\Delta x,\Delta x]\times\lbrack-\Delta p,\Delta p]$ in the
phase plane. The area of this rectangle is $4\Delta p\Delta x=4\hbar$ and its
largest inscribed ellipse $\Omega$ (\textquotedblleft John
ellipse\textquotedblright) is defined by the inequality
\[
\Omega:\frac{x^{2}}{4\Delta x^{2}}+\frac{p^{2}}{4\Delta p^{2}}\leq1
\]
and therefore has area $\pi\hbar$. This ellipse $\Omega$ is thus a
\textquotedblleft quantum blob\textquotedblright, \textit{i.e}. a smallest
phase plane unit compatible with the uncertainty principle of quantum
mechanics \cite{physletta,go09,blob,goluPR}. We \textit{postulate} that this
ellipse $\Omega$ is the covariance ellipsoid \cite{Birk,go09,Littlejohn} of
the quantum state of the particle under observation, in which case the
covariance matrix of the state is
\[
\Sigma=%
\begin{pmatrix}
\sigma_{xx} & \sigma_{xp}\\
\sigma_{px} & \sigma_{pp}%
\end{pmatrix}
=%
\begin{pmatrix}
2\Delta x^{2} & 0\\
0 & 2\Delta p^{2}%
\end{pmatrix}
\]
where $\sigma_{xx}$ and $\sigma_{pp}$ are the standard deviations; the
covariance $\sigma_{xp}=\sigma_{px}$ is zero. This postulate \textit{a
posteriori} identifies $\Delta x$ and $\Delta p$ with $\sqrt{\sigma_{xx}/2}$
and $\sqrt{\sigma_{pp}/2}$ and the relation $\Delta p\Delta x=\hbar$ is
equivalent to the saturated Heisenberg inequality $\sigma_{xx}\sigma
_{pp}=\frac{1}{4}\hbar^{2}$ and this is possible if and only if the state of
the particle is a Gaussian
\[
\psi(x)=\left(  \tfrac{1}{2\pi\sigma_{xx}}\right)  ^{1/4}e^{-\frac{x^{2}%
}{4\sigma_{xx}}}=\left(  \tfrac{1}{4\pi\Delta x^{2}}\right)  ^{1/4}%
e^{-x^{2}/8\Delta x^{2}}.
\]
Let us next turn to the more general case where the equality (\ref{equal1}) is
replaced with the inclusion
\begin{equation}
\lbrack-\Delta x,\Delta x]^{\hbar}=[-\hbar/\Delta x,\hbar/\Delta
x]\subset\lbrack-\Delta p,\Delta p] \label{equal2}%
\end{equation}
which we assume to be strict (i.e. $\Delta p\Delta x>\hbar$). The area
$4\Delta p\Delta x$ of the rectangle $[-\Delta x,\Delta x]\times\lbrack-\Delta
p,\Delta p]$ is here larger than $4\hbar$ and the John ellipse $(x^{2}/4\Delta
x^{2})+(p^{2}/4\Delta p^{2})\leq1$ has area $>\pi\hbar$ so it is not a quantum
blob and cannot therefore be identified with the covariance ellipse of a pure
quantum state. (but it will be the covariance ellipse of a mixed state, as we
will see in Section \ref{sec3}). What we do instead is to remark that in the
previously considered case $\Delta p\Delta x=\hbar$ the orthogonal projections
of the John ellipse on the $x$ and $p$ axes are precisely the intervals $X$
and $P=X^{\hbar}$. This suggests that we define as candidates for being
covariance ellipses those inscribed in $[-\Delta x,\Delta x]\times
\lbrack-\Delta p,\Delta p]$ having orthogonal projections $[-\Delta x,\Delta
x]$ and $[-\Delta p,\Delta p]$, this will work provided that these ellipses
are quantum blobs, that is that they have area $\pi\hbar$. Writing as before
the covariance matrix as $\Sigma=%
\begin{pmatrix}
\sigma_{xx} & \sigma_{xp}\\
\sigma_{px} & \sigma_{pp}%
\end{pmatrix}
$ its inverse is%
\[
\Sigma^{-1}=\frac{1}{D}%
\begin{pmatrix}
\sigma_{xx} & -\sigma_{xp}\\
-\sigma_{px} & \sigma_{pp}%
\end{pmatrix}
\text{ \ },\text{ \ }D=\sigma_{xx}\sigma_{pp}-\sigma_{xp}^{2}%
\]
so that the associated covariance ellipse $\Omega:\frac{1}{2}z^{T}\Sigma
^{-1}z\leq1$ is defined by the inequality%
\begin{equation}
\Omega:\dfrac{\sigma_{pp}}{2D}x^{2}-\frac{\sigma_{xp}}{D}px+\dfrac{\sigma
_{xx}}{2D}p^{2}\leq1 \label{2D}%
\end{equation}
where $\sigma_{xx}>0$, $\sigma_{pp}>0$, and $\sigma_{xp}$ (and hence $D$) have
to be determined from the knowledge of $\Delta x$ and $\Delta p$. The
orthogonal projections of $\Omega$ on the $x$ and $p$ axes are easily
calculated and one finds%
\begin{equation}
\Omega_{X}:(\dfrac{\sigma_{pp}}{2D}-\frac{\sigma_{xp}^{2}}{2D\sigma_{xx}%
})x^{2}\leq1\text{ },\text{ \ }\Omega_{P}:(\dfrac{\sigma_{xx}}{2D}%
-\frac{\sigma_{xp}^{2}}{2D\sigma_{pp}})p^{2}\leq1\text{ }%
\end{equation}
that is, simplifying, $\Omega_{X}:x^{2}/2\sigma_{xx}\leq1$ and $\Omega
_{P}:p^{2}/2\sigma_{pp}\leq1$. The projections are thus the intervals
\begin{equation}
\Omega_{X}=[-\sqrt{2\sigma_{xx}},\sqrt{2\sigma_{xx}}]\text{\ \textit{\ }%
}\mathit{,}\text{\textit{ \ } }\Omega_{P}=[-\sqrt{2\sigma_{pp}},\sqrt
{2\sigma_{pp}}\dot{]} \label{intervals}%
\end{equation}
and identifying them with $[-\Delta x,\Delta x]$ and $[-\Delta p,\Delta p]$
yields $\sigma_{xx}=\Delta x^{2}/2$ and $\sigma_{pp}=\Delta p^{2}/2$. The
inclusion (\ref{equal2}) being equivalent to $\Delta p\Delta x$ $>\hbar$ we
have $\sigma_{xx}\sigma_{pp}>\frac{1}{4}\hbar^{2}$ (strict Heisenberg
inequality). Since we are looking for a Gaussian pure state the
Robertson--Schr\"{o}dinger inequality must be saturated, which leads to
$D=\frac{1}{4}\hbar^{2}$ and the covariance $\sigma_{xp}$ can then be
calculated solving the equation $D=\sigma_{xx}\sigma_{pp}-\sigma_{xp}%
^{2}=\frac{1}{4}\hbar^{2}$, leading to two solutions $\sigma_{xp}=\pm
(\sigma_{xx}\sigma_{pp}-\frac{1}{4}\hbar^{2})^{1/2}$. The corresponding
Gaussian wave functions
\[
\psi^{\pm}(x)=\left(  \tfrac{1}{2\pi\sigma_{xx}}\right)  ^{1/4}e^{-\frac
{x^{2}}{4\sigma_{xx}}}e^{\pm\frac{i\sigma_{xp}}{2\hbar\sigma_{xx}}x^{2}}%
\]
are thus, in terms of $\Delta x^{2}$ and $\Delta p^{2}$,
\begin{align*}
\psi^{\pm}(x)  &  =\left(  \tfrac{1}{4\pi\Delta x^{2}}\right)  ^{1/4}%
e^{-x^{2}/8\Delta x^{2}}e^{\pm\frac{i\sigma_{xp}}{2\hbar\sigma_{xx}}x^{2}}\\
\sigma_{xp}  &  =\sqrt{2\Delta p^{2}-\hbar^{2}/8\Delta x^{2}}.
\end{align*}
These two wavefunctions are thus \textquotedblleft Pauli
partners\textquotedblright; the existence of a double solution is symptomatic
of the generic non-uniqueness of the solution of the Pauli problem
\ \cite{espo,gopauli,Weigert}.

\section{The multi-dimensional case\label{sec2}}

\subsection{Pure Gaussian states}

We now consider the case of a quantum system with $n$ degrees of freedom; its
configuration space $\mathbb{R}_{x}^{n}$ will be denoted $\ell_{X}$. The
corresponding momentum space $\ell_{P}=\mathbb{R}_{p}^{n}$ is as usual
identified with the dual space $(\mathbb{R}_{x}^{n})^{\ast}$ and the pairing
between momentum and position vectors will be loosely written as an inner
product $p\cdot x=p_{1}x_{1}%
\acute{}%
+\cdot\cdot\cdot+p_{n}x_{n}$. Assume, as in the example of Section \ref{sec1},
that after a large number of independent position measurements the quantum
system has been found to be contained in an ellipsoid $X(x_{0})=x_{0}+X$ with
$X:x^{T}Ax\leq\hbar$ where $A$ is a symmetric and positive definite real
$n\times n$ matrix. The polar dual of $X$ is $X^{\hbar}:p^{T}A^{-1}p\leq\hbar$
in $\ell_{P}$,, and we next consider the ellipsoid $P(p_{0})=p_{0}+X^{\hbar}$
centered at some value $p_{0}$ obtained by averaging over measured values of
omentum. We now identify the quantum system with the Cartesian product
$X(x_{0})\times(P(p_{0})$. This is a convex body in phase space $\ell
_{X}\times\ell_{P}=\mathbb{R}_{x}^{n}\times\mathbb{R}_{p}^{n}$, centered at
$z_{0}=(x_{0},p_{0})$, hence it might be a good idea to look for its John
ellipsoid \cite{Ball} $(X(x_{0})\times(P(p_{0}))_{\mathrm{John}}$,
\textit{i.e.} the unique ellipsoid of maximum volume contained in
$X(x_{0})\times P(p_{0})$. We begin with the easy case where $x_{0}$ and
$p_{0}$ are both zero, and $X$ is the ball $B_{X}^{n}(r):|x|\leq r$ in
$\ell_{X}$, in which case $X^{\hbar}=B_{X}^{n}(\hbar/r)$ \cite{ABMB,Vershynin}%
. Now, an easy geometric argument \cite{BSM} shows that
\begin{equation}
\left(  B_{X}^{n}(\sqrt{\hbar})\times B_{P}^{n}(\sqrt{\hbar})\right)
_{\mathrm{John}}=B^{2n}(\sqrt{\hbar}) \label{balljohn}%
\end{equation}
and a trivial rescaling by the symplectic mapping $S:(x,p)\longmapsto
(rx/\sqrt{\hbar},p\sqrt{\hbar}/r)$ shows that the John ellipsoid of $B_{X}%
^{n}(r)\times B_{X}^{n}(\hbar/r)$ is%
\[
\left(  B_{X}^{n}(r)\times B_{X}^{n}(\hbar/r)\right)  _{\mathrm{John}}%
:\frac{r^{2}}{\hbar}|x|^{2}+\frac{\hbar}{r^{2}}|p|^{2}\leq\hbar.
\]
Since this is the image of the ball $B^{2n}(\sqrt{\hbar})$ by a symplectic
mapping, $\left(  B_{X}^{n}(r)\times B_{X}^{n}(\hbar/r)\right)
_{\mathrm{John}}$ is a quantum blob, by the same argument as in Section
\ref{sec1}, this quantum blob is the covariance ellipsoid corresponding to the
$2n\times2n$ dimensional covariance matrix%
\[
\Sigma=%
\begin{pmatrix}
(\hbar^{2}/2r^{2})I_{n\times n} & 0_{n\times n}\\
0_{n\times n} & (r^{2}/2)I_{n\times n}%
\end{pmatrix}
\]
of a minimum uncertainty quantum state $|\psi\rangle$ with wave function the
squeezed Gaussian
\[
\psi(x)=\left(  \frac{1}{2\pi}\right)  ^{1/4}e^{-x^{2}/4\sigma_{xx}}.
\]
Let us turn to the most general pure Gaussian case, represented by a
wavefunction%
\begin{multline}
\psi(x)=\left(  \tfrac{1}{2\pi}\right)  ^{n/4}(\det\Sigma_{XX})^{-1/4}%
\label{Gauss3}\\
\times\exp\left[  -(x-x_{0})^{T}\left(  \frac{1}{4}\Sigma_{XX}^{-1}+\frac
{i}{2\hbar}\Sigma_{XP}\Sigma_{XX}^{-1}\right)  (x-x_{0})\right] \nonumber
\end{multline}
corresponding to a covariance matrix%
\[
\Sigma=%
\begin{pmatrix}
\Sigma_{xx} & \Sigma_{xp}\\
\Sigma_{px} & \Sigma_{pp}%
\end{pmatrix}
\text{ \ },\text{ \ }\Sigma_{xp}^{T}=\Sigma_{px}%
\]
where the block matrices $\Sigma_{xx}$ and $\Sigma_{pp}$
\[
\Sigma_{XX}=(\sigma_{x_{j}x_{k}})_{1\leq j,k\leq n}\text{ \ and \ }\Sigma
_{PP}=(\sigma_{p_{j}p_{k}})_{1\leq j,k\leq n}%
\]
are symmetric and positive definite and defined by%
\[
\sigma_{x_{j}x_{k}}=\int x_{j}x_{k}|\psi(x)|^{2}d^{n}x\text{ \ , \ }%
\sigma_{p_{j}p_{k}}=\int p_{j}p_{k}|\widehat{\psi}(p)|^{2}d^{n}p.
\]
The state being a pure Gaussian, the covariance matrix must satisfy the
saturated Robertson--Schr\"{o}dinger inequalities; in matrix form
\cite{gopauli}
\begin{equation}
\Sigma_{PP}\Sigma_{XX}-\Sigma_{XP}^{2}=\tfrac{1}{4}\hbar^{2}I_{n\times n}
\label{RSUP}%
\end{equation}
(this relation simply reflects the fact that $\frac{2}{\hbar}\Sigma$ is
symplectic since the covariance ellipsoid $\Omega:\frac{1}{2}z^{T}\Sigma
^{-1}z\leq1$ is a quantum blob \cite{blob}). We claim that this case
corresponds to the case where the observed set $x_{0}+X$ is an ellipsoid
centered at $x_{0}$. to simplify notation we assume $x_{0}=0$ and
$X:x^{T}Ax\leq\hbar$ \ where $A$ is a real symmetric and positive definite
$n\times n$ matrix. As before we have $X^{\hbar}:p^{T}A^{-1}p\leq\hbar$. Let
now $P$ be a centered ellipsoid of momentum space $\ell_{P}=\mathbb{R}_{p}%
^{n}$ containing $X^{\hbar}$ and consider the Cartesian product $X\times P$.
We claim that $X\times P$ contains at least one quantum blob. Let in fact
$P:p^{T}Ap\leq\hbar$ for a symmetric positive definite matrix $B$. The
inclusion $X^{\hbar}\subset P$ is equivalent to the matrix inequality $B\leq
A^{-1}$ ($\leq$ standing for the L\"{o}wner ordering of matrices); we will
write this condition simply as $AB\leq I_{n\times n}$ which means that the
eigenvalues of $AB$ must be smaller or equal to one. Let us now return to the
covariance matrix $\Sigma$; since $\frac{\hbar}{2}\Sigma^{-1}$ (and hence
$\frac{2}{\hbar}\Sigma$) is symplectic, the inverse $\Sigma^{-1}$ is easily
calculated and one finds
\begin{equation}
\Sigma^{-1}=\frac{4}{\hbar^{2}}%
\begin{pmatrix}
\Sigma_{PP} & -\Sigma_{PX}\\
-\Sigma_{XP} & \Sigma_{XX}%
\end{pmatrix}
. \label{sigmainv}%
\end{equation}
Using the theory of Schur complements as in \cite{BSM,FOOP}, one finds that
the orthogonal projections of the covariance ellipsoid $\Omega:\frac{1}%
{2}z^{T}\Sigma^{-1}z\leq1$ on, respectively, the configuration space $\ell
_{X}$ and the momentum space $\ell_{P}$ are the ellipsoids
\begin{align}
\Omega_{X}  &  :x^{T}(\Sigma_{PP}-\Sigma_{PX}\Sigma_{XX}^{-1}\Sigma_{XP}%
)x\leq\frac{\hbar^{2}}{2}\text{ }\label{ortho1}\\
\Omega_{P}  &  :p^{T}(\Sigma_{XX}-\Sigma_{XP}\Sigma_{PP}^{-1}\Sigma_{PX}%
)p\leq\frac{\hbar^{2}}{2}. \label{ortho2}%
\end{align}
Using the fact that $\frac{2}{\hbar}\Sigma$ \ is symplectic we have%
\begin{gather*}
\Sigma_{XX}\Sigma_{PP}-\Sigma_{XP}^{2}=\frac{\hbar^{2}}{4}I_{n\times n}\\
\Sigma_{XX}\Sigma_{PX}=\Sigma_{XP}\Sigma_{XX}\text{ },\text{ \ }\Sigma
_{PX}\Sigma_{PP}=\Sigma_{PP}\Sigma_{XP}%
\end{gather*}
so that
\begin{align*}
\Sigma_{PP}-\Sigma_{PX}\Sigma_{XX}^{-1}\Sigma_{XP}  &  =(\Sigma_{PP}%
\Sigma_{XX}-\Sigma_{PX}\Sigma_{XX}^{-1}\Sigma_{XP}\Sigma_{XX})\Sigma_{XX}%
^{-1}\\
&  =(\Sigma_{PP}\Sigma_{XX}-\Sigma_{PX}^{2})\Sigma_{XX}^{-1}\\
&  =\tfrac{1}{4}\hbar^{2}\Sigma_{XX}^{-1}%
\end{align*}
the last equality in view of (\ref{RSUP}); similarly%
\[
\Sigma_{XX}-\Sigma_{XP}\Sigma_{PP}^{-1}\Sigma_{PX}=\tfrac{1}{4}\hbar^{2}%
\Sigma_{PP}^{-1}.
\]
Summarizing, we arrive at the following simple form for the projections:%
\begin{equation}
\Omega_{X}:\frac{1}{2}x^{T}\Sigma_{XX}^{-1}x\leq1\text{ \ and }\Omega
_{P}:\frac{1}{2}p^{T}\Sigma_{PP}^{-1}p\leq1. \label{orthoproj}%
\end{equation}
The final step consists in identifying $\Omega_{X}$ and $\Omega_{P}$ with the
ellipsoids $X:x^{T}Ax\leq\hbar$ and $P:p^{T}Ap\leq\hbar$ defined above;
recalling that the condition $X^{\hbar}\subset P$ is equivalent to $AB\leq
I_{n\times n}$ this leads to the relations
\begin{equation}
\Sigma_{XX}=\frac{\hbar}{2}A^{-1}\text{ , }\Sigma_{PP}=\frac{\hbar}{2}B^{-1}
\label{sigxx}%
\end{equation}
and hence, in view of (\ref{RSUP}),
\begin{equation}
\Sigma_{XP}^{2}=\Sigma_{PP}\Sigma_{XX}-\tfrac{1}{4}\hbar^{2}I_{n\times n}.
\label{sigxp}%
\end{equation}
This equation has multiple solutions $\Sigma_{XP}$, reflecting again the
non-uniqueness of the solution to Pauli's problem

\subsection{Mixed Gaussian states}

We ask now whether we can obtain in a similar way the most general mixed
Gaussian quantum state, represented by its Wigner distribution%
\begin{equation}
W_{\rho}(x)=\left(  \frac{1}{2\pi}\right)  ^{n/2}\frac{1}{\sqrt{\det\Sigma}%
}\exp\left(  -\frac{1}{2}(x-x_{0})^{T}\Sigma^{-1}(x-x_{0})\right)  .
\label{wrho}%
\end{equation}
The covariance matrix
\[
\Sigma=%
\begin{pmatrix}
\Sigma_{xx} & \Sigma_{xp}\\
\Sigma_{px} & \Sigma_{pp}%
\end{pmatrix}
\]
must satisfy the well-known quantum condition \cite{dutta,Birk,go09,blob}
\begin{equation}
\Sigma+\frac{i\hbar}{2}J\geq0 \label{quantum}%
\end{equation}
where $J=%
\begin{pmatrix}
0_{n\times n} & I_{n\times n}\\
-I_{n\times n} & 0_{n\times n}%
\end{pmatrix}
$ is the standard symplectic matrix; it means that the eigenvalues of the
Hermitian matrix $\Sigma+\frac{i\hbar}{2}J$ are all $\geq0$ (and hence
$\Sigma>0$). Condition (\ref{quantum}) is just the Robertson--Schr\"{o}dinger
uncertainty principle
\begin{equation}
\sigma_{x_{j}x_{j}}\sigma_{p_{j}p_{j}}-\sigma_{x_{j}p_{j}}\geq\tfrac{1}%
{4}\hbar^{2}\text{ \ , \ }1\leq j\leq n \label{RS}%
\end{equation}
written in compact form. As we have shown in \cite{go09,blob,goluPR} condition
(\ref{quantum}) is equivalent to the following geometric condition:

\begin{quote}
The covariance ellipsoid $\Omega:\frac{1}{2}z^{T}\Sigma^{-1}z\leq1$ contains a
quantum blob $S(B^{2n}(\sqrt{\hbar})$ ($S\in\operatorname*{Sp}(n)$).
\end{quote}

Assuming for notational simplicity that $x_{0}=0$ and $p_{0}=0$ we consider,
as before, centered ellipsoids $X:x^{T}Ax\leq\hbar$ and $P:p^{T}Ap\leq\hbar$
in $\ell_{X}$ and $\ell_{P}$ satisfying $X^{\hbar}\subset P$ (recall that this
inclusion is equivalent to the matrix inequality $AB\leq I_{n\times n}$). We
have already studied the case $X^{\hbar}=P$ in last section (pure states) so
we assume that the inclusion $X^{\hbar}\subset P$ is strict: $X^{\hbar
}\subsetneqq P$. Let us denote by $\Omega_{\mathrm{John}}$ the John ellipsoid
\cite{Ball} of $X\times P$: it is by definition the unique ellipsoid of
maximal volume contained in $X\times P$. By elementary symmetry conditions
$\Omega_{\mathrm{John}}$ is centered at the origin $0$ of phase space; we
claim that, in addition, $\Omega_{\mathrm{John}}$ contains a quantum blob
$S(B^{2n}(\sqrt{\hbar})$. Here is why. We have $X=A^{-1/2}(B_{X}^{n}%
(\sqrt{\hbar}))$ and $P=B^{-1/2}(B_{P}^{n}(\sqrt{\hbar}))$ and hence%
\[
X\times P=%
\begin{pmatrix}
A^{-1/2} & 0_{n\times n}\\
0_{n\times n} & B^{-1/2}%
\end{pmatrix}
(B_{X}^{n}(\sqrt{\hbar})\times B_{P}^{n}(\sqrt{\hbar}));
\]
using the covariance of the John ellipsoid under linear transformations and
taking into account formula (\ref{balljohn}) it follows that%
\[
(X\times P)_{\mathrm{John}}=M(B^{2n}(\sqrt{\hbar}))
\]
where $M$ is the matrix
\[%
\begin{pmatrix}
A^{-1/2} & 0_{n\times n}\\
0_{n\times n} & B^{-1/2}%
\end{pmatrix}
=%
\begin{pmatrix}
I_{n\times n} & 0_{n\times n}\\
0_{n\times n} & (AB)^{-1/2}%
\end{pmatrix}%
\begin{pmatrix}
A^{-1/2} & 0\\
0 & A^{1/2}%
\end{pmatrix}
.
\]
Noting that $S_{A^{1/2}}=%
\begin{pmatrix}
A^{-1/2} & 0\\
0 & A^{1/2}%
\end{pmatrix}
$ is a symplectic matrix the ellipsoid $S_{A^{1/2}}(B^{2n}(\sqrt{\hbar}))$ is
a quantum blob; the claim follows since $S_{A^{1/2}}(B^{2n}(\sqrt{\hbar
}))\subset(X\times P)_{\mathrm{John}}$ because $(AB)^{-1/2}\geq I_{n\times n}$
implies that
\[
S_{A^{1/2}}(B^{2n}(\sqrt{\hbar}))\subset M(B^{2n}(\sqrt{\hbar}))=(X\times
P)_{\mathrm{John}}.
\]
The quantum states (\ref{wrho}) are now determined by the covariance matrix
$\Sigma$ defined as being the unique symmetric positive definite matrix such
that $\Omega_{\mathrm{John}}:\frac{1}{2}z^{T}\Sigma^{-1}z\leq1$; that the
quantum condition (\ref{quantum}) holds is clear since $_{\mathrm{John}}$
contains a quantum blob.

\section{Symplectic action on mixed Gaussian states\label{sec3}}

\section{Discussion}

While our constructions are mathematically rigorous, their physical
interpretation requires the postulate that if position measurements on a
physical system are contained in a configuration space ellipsoid $x_{0}+X$
then the momentum measurements must be contained in a momentum space ellipsoid
$p_{0}+P$ satisfying the polarity condition $X^{\hbar}\subset P$. While this
postulate is certainly plausible, it has to be verified by experimental
measurements or simulations. Also, the hypothesis that a set of ideally
precise measurements should be modeled by an ellipsoid is plausible, but rater
\textit{ad hoc}. It can however be justified by standard techniques of convex
optimization: suppose we have performed a large number $N$ of position
measurements, leading to a set $X=\{x_{1},...,x_{N}\}$ of points. Taking the
convex hull of this set we get convex polyhedron $X_{\mathrm{conv}}$ in
configuration space, and the latter can be approximated by the John ellipsoid
$X_{\mathrm{John}}$ of that polyhedron. Now, taking polar duals transforms the
John ellipsoid of a convex set into the L\"{o}wner ellipsoid of the dual of
this set \cite{ABMB,Vershynin} (the L\"{o}wner ellipsoid of a convex set is
the unique ellipsoid with smallest volume containing that set \cite{Ball}). In
this way we get an approximation to the quantum state; how good this
approximation is will not be discussed here; it requires rather standard
techniques from convex optimization and should not be overwhelmingly
difficult. The aim of this work is to advertise the fact that we have a method
for associating to ellipsoids in configuration space a host of well-defined
Gaussian sates. Our method can (and will) be extended to general convex
subsets of configuration space; the difficulty lies then in the fact that the
polar dual of an arbitrary has to be defined using the so-called Santal\'{o}
point of the set; it is the point which minimizes the Mahler volume of the set
in question, and this related to the uncertainty principle as we have shortly
discussed in \cite{FOOP,MC}: the structure involved is complicated but very
rich and certainly deserves to be studied further.

\section*{APPENDIX\ A: Symplectic Matrices}

We collect here some elementary facts from the theory of symplectic matrices;
for details and proofs see \cite{Birk}. The standard symplectic form $\omega$
on $\mathbb{R}_{z}^{2n}\equiv\mathbb{R}_{x}^{n}\times\mathbb{R}_{p}^{n}$ can
be written in matrix form as
\[
\omega(z,z^{\prime})=(z^{\prime})^{T}Jz=Jz\cdot z^{\prime}%
\]
where $J$ is the standard symplectic matrix. By definition:%
\[
J=%
\begin{pmatrix}
0_{n\times n} & I_{n\times n}\\
-I_{n\times n} & 0_{n\times n}%
\end{pmatrix}
.
\]
The associated symplectic group $\operatorname*{Sp}(n)$ consists of all real
$2n\times2n$ matrices $S$ such that $S^{T}JS=J$ or, equivalently, $SJS^{T}=J$.
These relations imply \cite{Birk} that if we write $S$ in the block form
\begin{equation}
S=%
\begin{pmatrix}
A & B\\
C & D
\end{pmatrix}
\label{ABCD}%
\end{equation}
the $S$ is symplectic if and only if the $n\times n$ blocks $A,B,C,D$ satisfy
the sets of equivalent conditions
\begin{align}
A^{T}C\text{, }B^{T}D\text{ \ \textit{symmetric, and} }A^{T}D-C^{T}B  &
=I_{n\times n}\label{cond1}\\
AB^{T}\text{, }CD^{T}\text{ \ \textit{symmetric, and} }AD^{T}-BC^{T}  &
=I_{n\times\times n}. \label{cond2}%
\end{align}
It follows that the inverse of $S\in\operatorname*{Sp}(n)$ has the simple form%
\begin{equation}
S^{-1}=%
\begin{pmatrix}
D^{T} & -B^{T}\\
-C^{T} & A^{T}%
\end{pmatrix}
. \label{inverse}%
\end{equation}
A symplectic automorphism $U$ is called a \textit{symplectic}
\textit{rotation} if $U\in\operatorname*{Sp}(n)\cap O(2n,\mathbb{R})$ where
$O(2n,\mathbb{R})$ is the usual orthogonal group. In the case $n=1$ this is
just the usual rotation group $SO(2n,\mathbb{R})$. We denote by $U(n)$ the
group of all symplectic rotations; one shows \cite{Birk} that $U(n)$ is the
image in $\operatorname*{Sp}(n)$ of the complex unitary group $U(n,\mathbb{C}%
)$ by the embedding%
\[
\iota:A+iB\longmapsto%
\begin{pmatrix}
A & B\\
-B & A
\end{pmatrix}
.
\]
A matrix $%
\begin{pmatrix}
A & B\\
-B & A
\end{pmatrix}
$ is thus a symplectic rotation if and only if the blocks $A$ and $B$ satisfy
the conditions%
\begin{align}
A^{T}B  &  =B^{T}A\text{ \ and }A^{T}A+B^{T}B=I_{n\times n}\label{uni1}\\
AB^{T}  &  =BA^{T}\text{ \ and }AA^{T}+BB^{T}=I_{n\times n} \label{uni2}%
\end{align}

\section*{APPENDIX\ B: Polar Duality}

See \cite{FOOP} for a short account of polar duality. The treatises
\cite{ABMB,Vershynin} give detailed accounts of the theory in the context of
convex geometry.

Let $X$ be a convex body in configuration space $\mathbb{R}_{x}^{n}$ (a convex
body is a compact convex set with non-empty interior). We assume in addition
that $X$ contains $0$ in its interior. This is the case if, for instance, $X$
is symmetric: $X=-X$. The \emph{polar dual} of $X$ is the subset
\begin{equation}
X^{\hbar}=\{p\in\mathbb{R}_{x}^{n}:\sup\nolimits_{x\in X}(p\cdot x)\leq\hbar\}
\label{omo}%
\end{equation}
of the dual space $\mathbb{R}_{p}^{n}\equiv(\mathbb{R}_{x}^{n})^{\ast}$.
Notice that it trivially follows from the definition that $X^{\hbar}$ is
convex and contains $0$ in its interior. In the mathematical literature one
usually chooses $\hbar=1$, in which case one writes $X^{o}$ for the polar
dual; we have $X^{\hbar}=\hbar X^{o}$. The following properties are straightforward:

\begin{center}%
\begin{tabular}
[c]{|l|l|l|}\hline
\textit{Reflexivity (bipolarity)}: & $(X^{\hbar})^{\hbar}=X$ & P1\\\hline
\textit{Antimonotonicity: } & $X\subset Y\Longrightarrow Y^{\hbar}\subset
X^{\hbar}$ & P2\\\hline
\textit{Scaling property} & $A\in GL(n,\mathbb{R})\Longrightarrow(AX)^{\hbar
}=(A^{T})^{-1}X^{\hbar}$. & P3\\\hline
\end{tabular}

\end{center}

The following elementary properties of polar duality hold:

\textit{(i)} Let $B_{X}^{n}(R)$ (\textit{resp}. $B_{P}^{n}(R)$) be the ball
$\{x:|x|\leq R\}$ in $\mathbb{R}_{x}^{n}$ (\textit{resp}. $\{p:|p|\leq R\}$ in
$\mathbb{R}_{p}^{n}$). Then
\begin{equation}
B_{X}^{n}(R)^{\hbar}=B_{P}^{n}(\hbar/R)~. \label{BhR}%
\end{equation}
In particular
\begin{equation}
B_{X}^{n}(\sqrt{\hbar})^{\hbar}=B_{P}^{n}(\sqrt{\hbar}). \label{bhh}%
\end{equation}

\textit{(ii)} Let $A$ be a real invertible and symmetric $n\times n$ matrix
and $R>0$. The polar dual of the ellipsoid defined by $x^{T}Ax\leq R^{2}$ is
given by
\begin{equation}
\{x:x^{T}Ax\leq R^{2}\}^{\hbar}=\{p:p^{T}A^{-1}p\leq(\hbar/R)^{2}\}
\label{dualell}%
\end{equation}
and hence%
\begin{equation}
\{x:x^{T}Ax\leq\hbar\}^{\hbar}=\{p:p^{T}A^{-1}p\leq\hbar\}~. \label{dualellh}%
\end{equation}

\begin{acknowledgement}
Maurice de Gosson has been financed by the Grant P 33447 N of the Austrian
Science Fund FWF (Fonds zur F\"{o}rderung der wissenschaftlichen Forschung..
\end{acknowledgement}

\end{document}